\documentclass[preprint,epsfig,floats,aps]{revtex4}
\usepackage{epsfig}
\begin{document}
%\baselinestretch{5}
%\tightenlines
\topmargin=-0.3cm
\title{Net charge fluctuation and string fragmentation}
%\vspace{0.1in}
\author{Ben-Hao Sa$^{1,2,4,5,}$\footnote{Email: sabh@iris.ciae.ac.cn},
 Xu Cai$^{4,2}$, An Tai$^3$, Zhong-Qi Wang$^1$ and Dai-Mei Zhou$^2$}
\affiliation{
$^1$  China Institute of Atomic Energy, P. O. Box 275 (18),
      Beijing, 102413 China \\
$^2$  Institute of Particle Physics, Huazhong Normal University,
      Wuhan, 430079 China\\
$^3$  Department of Physics and Astronomy, University of California,
      at Los Angeles, Los Angeles, California 90095 \\
$^4$  CCAST (World Lab.), P. O. Box 8730 Beijing, 100080 China\\
$^5$  Institute of Theoretical Physics, Academia Sinica, Beijing,
      100080 China
}
%\maketitle
\begin{abstract}
We present simulation results of net charge fluctuation in $Au+Au$ collisions 
at $\sqrt{s_{nn}}$=130 GeV from a dynamic model, JPCIAE. The calculations are 
done for the quark-gluon phase before hadronization, the pion gas, the 
resonance pion gas from $\rho$ and $\omega$ decays and so on. The simulations 
of the charge fluctuation show that the discrepancy exists between the dynamic 
model and the thermal model for a pion gas and a resonance pion gas from 
$\rho$ and $\omega$ decays while the simulated charge fluctuation of the 
quark-gluon phase is close to the thermal model prediction. JPCIAE results of 
net charge fluctuation in the hardonic phase are nearly $4-5$ times larger 
than one for the quark-gluon phase, which implies that the charge fluctuation 
in the quark-gluon phase may not survive the hadronization (string 
fragmentation) as implemented in JPCIAE.
\end{abstract}
%\vspace{0.1in}
\maketitle

\begin{figure}[ht]
\centerline{\hspace{-0.5in}
\epsfig{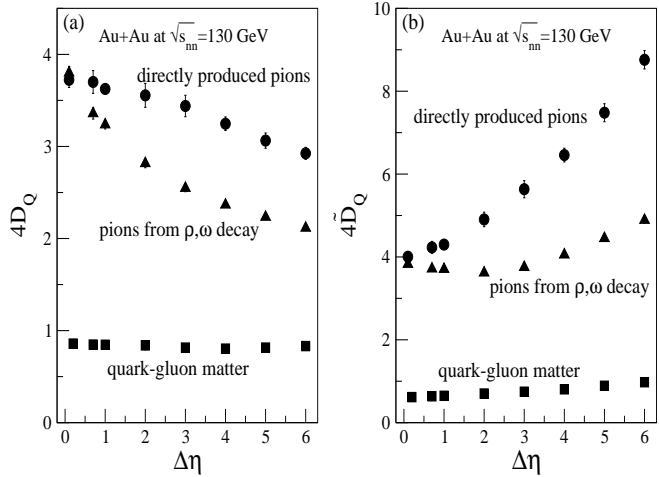}}
\hspace{0.25cm}
\caption{4$D_Q$ (left panel) and 4$\tilde{D_Q}$ (right panel) as a function 
of $\Delta \eta$ in $Au+Au$ collisions at $\sqrt{s_{nn}}$=130 GeV from JPCIAE 
simulations for the quark-gluon phase, pion gas (denoted by "directly produced 
pion"), and resonance pion gas from $\rho$ and $\omega$ decays.}
\label{flu_jp1}
\end{figure}

\begin{figure}[ht]
\centerline{\hspace{-0.5in}
\epsfig{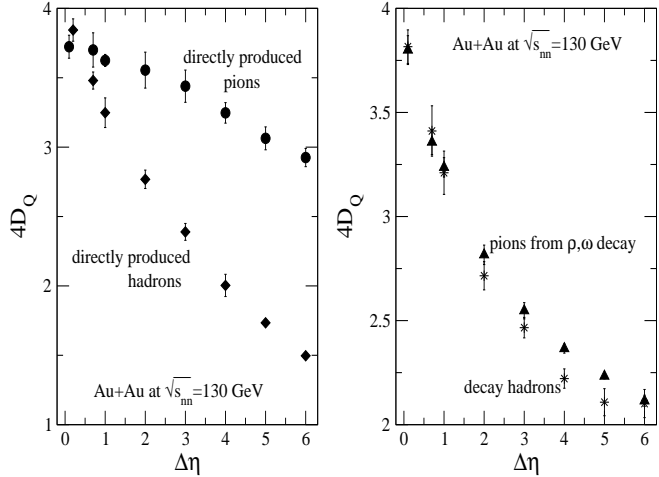}}
\hspace{0.25cm}
\caption{Comparison between pion net charge fluctuation and hadron net charge 
fluctuation in $Au+Au$ collisions at $\sqrt{s_{nn}}$=130 GeV: the left panel 
compares the results of "directly produced pions" with "directly produced 
hadrons", the right panel compares the results of "pions from $\rho$, $\omega$ 
decay" with "decay hadrons".}
\label{flu_jp2}
\end{figure}

\section{Introduction}
Study of Event-by-Event fluctuation reveals dynamics in relativistic heavy-ion
collisions. With the increase of particle multiplicity statistically 
significant measurements of E-by-E fluctuations became possible for the first 
time in $Pb+Pb$ collisions at 158A GeV/c \cite{rol,afa0,app,afa} and recently 
in $Au+Au$ collisions at $\sqrt{s_{nn}}$ =130 GeV \cite{phe,star}. Many 
theoretical studies based on hadronic transport models 
\cite{ble,liu,cap,bop,sa2} and effective models \cite{vol,bay,hei} were 
carried out trying to understand the effects of different dynamic processes on 
the E-by-E fluctuations. However, experimental data show that non-statistical 
contributions to the E-by-E fluctuation of average transverse momentum, 
$k/\pi$ ratio, and net charge multiplicity are small \cite{app,afa,phe}.

The charged particle ratio fluctuation was recently proposed as a signal of 
QGP formation \cite{jeo,asa}. The thermal model predicts that the magnitude of 
net charge fluctuation is $\sim$4 for a pion gas, $\sim$3 for a resonance pion 
gas (pion from $\rho$ and $\omega$ decays), and $\sim$0.75 for massless 
noninteracting quarks and gluons because the unit of charge in the QGP phase 
is 1/3 while it is 1 in the hadronic phase. Therefore, if the initial 
fluctuation survives hadronization, a measurement of the charge fluctuation 
would be able to tell whether a QGP phase is formed in the early stage of 
relativistic heavy-ion collisions. A review on E-by-E fluctuation can be found 
in \cite{sum}.

A crucial question here is how hadronization affects the charge fluctuation. 
Hadronization belongs to the non-perturbative regime and can not be solved 
from a first principle theory. The best knowledge of our understanding about 
hadronization has been implemented into two phenomenological models, string 
fragmentation \cite{lund} and cluster model \cite{clu}. In the string 
fragmentation, the parton cascade processes cease when the transverse momenta 
of emitted partons (ordered by their $p_{T}$) become smaller than a given cut, 
$p_{T}^{min}\sim$ 1 GeV/c. After that, a string is formed with a color triplet 
quark and a color anti-triplet antiquark (or diquark and anti-diquark) on each 
end of the string and gluons from the parton cascade distributed as `kinks' 
along the string. The formed string fragments into hadrons through quark-
antiquark pairs (diquark and anti-diquark pairs) production from the QCD 
vacuum. In the cluster model, at the end of the perturbative phase of parton 
cascade evolution, each gluon is forcibly splite into a quark-antiquark 
pair. Color singlet clusters are formed from the final quark-antiquark pairs 
(distinct from those in the gluon splitting), which have a minimal separation 
in coordinate and momentum space. The clusters subsequently decay 
independently into hadrons.

\begin{figure}[ht]
\centerline{\hspace{-0.5in}
\epsfig{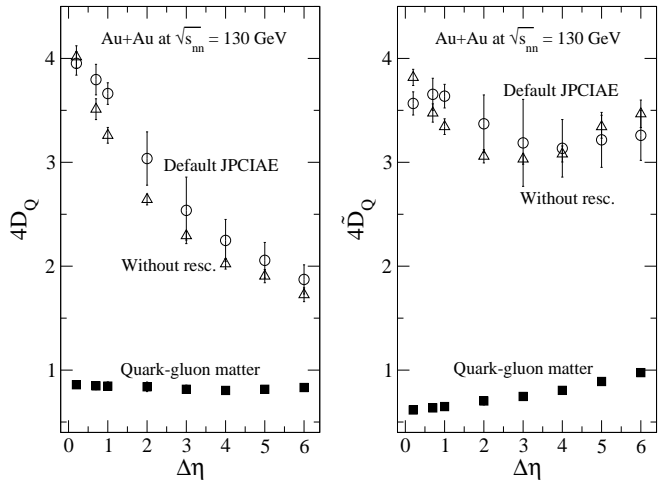}}
\hspace{0.25cm}
\caption{4$D_Q$ (left panel) and 4$\tilde{D_Q}$ (right panel) as a function of 
$\Delta \eta$ from JPCIAE calculations for $Au+Au$ collisions at $\sqrt{s_{nn}
}$=130 GeV. Open circles, open triangles and full squares are results of 
default JPCIAE, JPCIAE without rescattering and "quark-gluon phase", 
respectively.}
\label{flu_jp3}
\end{figure}

In elementary collisions, like in $e^{+}e^{-}$ and $pp$, the string is a well
-established object due to confinement force between quarks. However, in AA 
collisions, if a deconfined state is formed in the early stage of a collision, 
it is conceivable that the concept of string may become irrelevant any more 
since quarks are no longer bound together by the string-like color force. In 
this case, a coalescence picture of hadronization in the cluster model seems 
more applicable.

The net charge fluctuation was investigated using Monte-Carlo generators based 
on both string and cluster fragmentation in \cite{qinghui}. It was found that 
VNIb generator \cite{vnib}, which is based on cluster fragmentation, yields a 
net charge fluctuation close to that for a thermal QGP gas, a factor of 2 
smaller than that from the generators based on string fragmentation. It was 
argued that the difference may be due to the higher gluon density in VNIb. 
However, another possibility is that the net charge fluctuation in the parton 
phase can not survive string fragmentation, but can survive coalescence type 
of hadronization. In order to test this possibility one needs to compare net 
charge fluctuations calculated before string fragmentation in the partonic 
phase and after string fragmentation in the hadronic phase. In addition, the 
net charge fluctuations may also be affected by the decays of resonances and 
by rescatterings among hadrons produced during hadronization. These can be 
investigated in a dynamic model, JPCIAE.

In this paper a dynamic simulation based on the JPCIAE model was used to
study the charge fluctuation in $Au+A$u collisions at $\sqrt{s_{nn}}$=130 
GeV. In this calculation, the net charge fluctuations are calculated for the 
quark-gluon phase before parton fragmentation and for directly-produced pions 
and pions from $\rho$ and $\omega$ decays in the hadronic phase after string  
fragmentation. The results of this study are compared with the corresponding 
predictions of the thermal model. The simulations show that although the 
simulated net charge fluctuation of quark-gluon phase is close to that from 
the thermal model discrepancies are seen between the dynamic simulations and 
the thermal model predictions for pion gas and resonance pion gas from $\rho$ 
and $\omega$ decays. It is found that the net charge fluctuation in the 
hadronic phase from JPCIAE is nearly a factor of $4-5$ larger than one for 
the quark-gluon phase. The results indicate that the charge fluctuation in the 
partonic phase may not survive the hadronization (string fragmentation) as 
implemented in JPCIAE.

\section{Charge Fluctuations}
The deviation of a physical variable $x$ from its average value per event 
$<x>$ and its variance are defined as \cite{rei}
 \begin{equation}
 \delta x=x-<x>,
 \end{equation}
and
 \begin{equation}
 <(\delta x)^2>=<x^2>-<x>^2,
 \end{equation}
respectively. Suppose $\displaystyle{x\equiv R=N_+/N_-}$ to be the ratio of 
positively to negatively charged particle multiplicity, the corresponding 
variance is then 
 \begin{equation}
 <(\delta R)^2>=<R^2>-<R>^2.
 \end{equation}
Similarly, the variance of net charge multiplicity ($\displaystyle{Q=N_+-N_-}
$) reads, 
 \begin{equation}
 <(\delta Q)^2>=<Q^2>-<Q>^2,
 \end{equation}

However, what are used to study charge fluctuation in literature is 
 \begin{equation}
 D_R\equiv<N_{ch}><(\delta R)^2>,
 \end{equation}
or
 \begin{equation}
 D_Q\equiv\frac{<(\delta Q)^2>}{<N_{ch}>},
 \end{equation}
In above equations $\displaystyle{N_{ch}=N_++N_-}$ refers to the total charge 
multiplicity. When\\ $<N_{ch}> \gg <Q> $, a relation 
follows approximately \cite{jeo}
 \begin{equation}
 D_R\simeq4D_Q.
 \label{dd}
 \end{equation}

Two corrections have to be made in order to compare experimental data or 
dynamic simulations with the thermal model predictions. One correction, $C_y$, 
is done for the finite rapidity bin size and another correction, $C_{\mu}$, is 
for finite net charge. $C_{y}$ and $C_{\mu}$ are introduced in \cite{urqmd} as 
 \begin{eqnarray}
 C_y=1-\frac{<N_{ch}>_{\Delta y}}{<N_{ch}>_{total}}.\\
 C_{\mu}=\frac{<N_+>_{\Delta y}^2}{<N_->_{\Delta y}^2}.
 \end{eqnarray}

The corrected net charge fluctuation is denoted by 
 \begin{equation}
 \tilde{D_Q}=\frac{D_Q}{C_yC_{\mu}},
 \end{equation}
for instance.

\section{Model Calculations and Discussion}
The JPCIAE model is a transport model for AA collisions in which each hadron-
hadron collision, when its center-of-mass energy is larger than a given cut, 
is carried out by PYTHIA \cite{sjo1}, otherwise by conventional two-body 
interactions \cite{cugn,bert,tai1}. In PYTHIA, a string is stretched out after 
a collision with quarks and gluons being distributed alone the string, and the 
string later fragments into hadrons based on the Lund string fragmentation 
scheme. The gluons are produced through hard QCD scattering and bremsstrahlung
radiation. After one hadron-hadron collision, a formation time is given to 
produced hadrons. A hadron is only allowed to collide with other hadrons after 
it is formed. The hadron-hadron collisions will go on until no more collision 
would take place in the system. In the current version of PYTHIA produced 
partons do not rescatter with each other, unlike produced hadrons. Because 
both the partonic and hadronic phase are included in PYTHIA, it provides us a 
good dynamic tool to study the charge fluctuation for quark-gluon phase, pion 
gas and resonance pion gas from $\rho$ and $\omega$ decays. More importantly, 
we could use JPCIAE to investigate whether the charge fluctuation in the 
partonic phase can survive the hadronization (string fragmentation) processes. 
We refer to \cite{sa1} for more details about the JPCIAE model.

In line with the thermal model calculations, several dynamic simulations were 
performed first for 10\% most central $Au+Au$ collisions at $\sqrt{s_{nn}}$=
130 GeV based on the JPCIAE model for quark-gluon phase, the pion gas, and the 
resonance pion gas from $\rho$ and $\omega$ decays. The calculation of charge 
fluctuation for the quark-gluon phase is done before hadronization (string 
fragmentation), in which diquarks (anti-diquarks) are split into quarks 
(antiquarks) randomly, and the fractional charge of a quark (antiquark) is 
counted in its multiplicity and the gluon contribution in charge multiplicity 
is assumed to be 2/3 \cite{jeo}. For the pion gas, only directly produced 
pions from string fragmentation are counted in the calculation of charge 
fluctuation, in contrast to the pion gas from $\rho$, $\omega$ decays.

Fig. 1 gives the simulated net charge fluctuation in Au+Au collisions at 
$\sqrt{s_{nn}}$=130 GeV as a function of $\Delta\eta$ for 4$D_Q$ (left panel) 
and for 4$\tilde{D_Q}$ (right panel). The squares, circles, and triangles in 
this figure are the results for quark-gluon phase, pion gas (denoted by 
"directly produced pions"), and the pion gas from $\rho$, $\omega$ decays, 
respectively. One sees in this figure that the charge fluctuation from the 
dynamic simulation is very close to the thermal model prediction for the 
quark-gluon phase, but our result for the pion gas from $\rho$, $\omega$ 
decays is higher than the thermal model calculation. The JPCIAE result for 
the directly produced pions is consistent with the thermal model calculation 
only in the region of small $\Delta\eta$ and it increases rapidly with 
increasing $\Delta\eta$.

In the left panel of Fig. 2 the net charge fluctuation of "directly produced 
pions" are compared with the result of "directly produced hadrons" from 
JPCIAE. One sees that the charge fluctuation for "directly produced hadrons" 
is smaller than that for "directly produced pions", which is due to the 
constraint of charge conservation in string fragmentation when all hadrons are 
included. The result of "pions from $\rho$, $\omega$ decay" is compared, in 
the right panel of Fig. 2, with the result of "decay hadrons" (decay products 
from all unstable hadrons). One observes here that both results are close to 
each other, indicating that the correlation between positively and negatively 
charged hadrons from all unstable hadrons is similar.  

The net charge fluctuation as a function of $\Delta\eta$ from the JPCIAE 
simulations with (open circles) and without rescattering (open triangles) 
for $Au+Au$ collisions at $\sqrt{s_{nn}}$=130 GeV is compared with each other 
in Fig.(\ref{flu_jp3}). The JPCIAE result for the quark-gluon phase (solid 
squares) is also plotted for comparison. The results show that the 
rescattering effect on charge fluctuation is small. It is also shown in the 
right panel of Fig.(\ref{flu_jp3}) that the default JPCIAE results are nearly 
a factor of 4$-$5 larger than that for the quark-gluon phase.
               
In summary, we have performed dynamic simulations based on JPCIAE model for 
the quark-gluon phase, the pion gas, and the resonance pion gas from $\rho$ 
and $\omega$ decays for 10\% most central $Au+Au$ collisions at $\sqrt{s_{nn}}
$=130 GeV. Our results indicate that although the simulated net charge 
fluctuation of the quark-gluon phase is close to the thermal model prediction 
discrepancies are found between the dynamic simulation and the thermal model 
for the pion gas and resonance pion gas from $\rho$ and $\omega$ decays. It is 
also found that JPCIAE results of net charge fluctuation are nearly a factor 
of $4-5$ larger than that from the simulation for the quark-gluon phase. Our 
results thus indicate the charge fluctuation does not survive hadronization 
(string fragmentation) as implemented in JPCIAE.

Finally, the financial supports from NSFC (10135030 and 10075035) in China 
and DOE in USA are acknowledged.


\begin{thebibliography}{99}
\bibitem{rol}
G. Roland, in Proc. Workshop, QCD phase transitions, January 1997, Hirscheeg,
Austria.
\bibitem{afa0}
S. V. Afanasiev et al., NA49 Collaboration, hep-ex/0009053
\bibitem{app}
H. Appelsh\"{a}user et al., NA49 Collaboration, Phys. Lett. B 459, 679 (1999).
\bibitem{afa}
S. V. Afanasiev et al., NA49 Collaboration, Phys. Rev. Lett. 86, 1965 (2001).
\bibitem{phe}
K. Adcox, et al., PHENIX Collaboration, nucl-ex/0203014.
\bibitem{star}
Lanny Ray, STAR Collaboration, QM 2002, Nantes, France, July 18-24, 2002.
\bibitem{ble}
M. Bleicher, M. Belkacem, C. Ernst, H. Weber, L. Gerland, C. Spieles,
S. A. Bass, H. St\"{o}cker and W. Greiner, Phys. Lett. B 435, 9 (1998).
\bibitem{liu}
F. Liu, A. Tai, M. Ga\'{z}dzicki and R. Stock, Eur. Phys. J. C 8, 649 (1999).
\bibitem{cap}
A. Capella, E. G. Ferreiro and A. B. Kaidalov, hep-ph/9903338.
\bibitem{bop}
F. W. Bopp and J. Ranft, Eur. Phys. J. C 22, 171 (2001).
\bibitem{sa2}
Ben-Hao Sa, Xu Cai, An Tai, and Dai-Mei Zhou,  Phys. Rev. C 66, 044902 (2002).
\bibitem{vol}
S. A. Voloshin, V. Koch and H. G. Rither, Phys. Rev. C 60, 024901 (1999).
\bibitem{bay}
G. Baym and H. Heiselberg, Phys. Lett. B 469, 7 (1999).
\bibitem{hei}
H. Heiselberg and A. D. Jackson, Phys. Rev. C 63,  064904 (2001).
\bibitem{jeo}
S. Jeon and V. Koch, Phys. Rev. Lett. 83, 5435 (1999);
S. Jeon and V. Koch, Phys. Rev. Lett. 85, 2076 (2000).
\bibitem{asa}
M. Asakawa, U. Heinz and B. M\"{u}ller, Phys. Rev. Lett. 85, 2072 (2000).
\bibitem{sum} 
H. Heiselberg, Phys. Rep. 351, 161 (2001);\\
S. Jeon and V. Koch, hep-ph/0304012.
\bibitem{lund} 
B. Andersson, G. Gustafson, G. Ingelman and T. Sj\"ostrand, Phys. Rep. 97, 31 
(1983).
\bibitem{clu} 
G. Marchesini and B. R. Webber, Nucl. Phys. B238, 1 (1984).
\bibitem{qinghui} 
Q. H. Zhang, V. Topor Pop, S. Jeon and C. Gale, Phys. ReV. C66, 014909 (2002).
\bibitem{vnib} 
K. Geiger, Phys. Rep. 258, 237 (1995).
\bibitem{rei}
L. E. Reichl, A modern course in statistical physics, University of Texas
press, Austin, 1980;\\
R. Kubo, Statistical mechanics, North-Holland Publishing Company-Amsterdam,
1965.
\bibitem{urqmd} 
M. Bleicher, S. Jean and V. Koch, Phys. Rev. C 62, 061902(R) (2000).
\bibitem{sjo1}
T. Sj$\ddot{o}$strand, Comput. Phys. Commun. 82, 74 (1994).
\bibitem{cugn}
J. Cugnon, T.  Mizutani, and J. Vandermeulen, Nucl. Phys. A 352, 505 (1981).
\bibitem{bert}
G. F. Bertsch and S. Das Gupta, Phys. Rep. 160, 189 (1988);\\
A. Bonasera, F. Gulminelli, and J. Molitoris, Phys. Rep. 243, 1(1994).
\bibitem{tai1}
An Tai and Ben-Hao Sa, Comput. Phys. Commun. 116, 353 (1999).
\bibitem{sa1}
Ben-Hao Sa, An Tai, Hui Wang and Feng-He Liu, Phys. Rev. C 59, 2728 (1999);\\
Ben-Hao Sa and An Tai, Phys. Rev. C 62, 044905 (2000).
\end{thebibliography}
\end{document}